\pdfoutput=1
\documentclass[journal]{IEEEtran}

\usepackage{amsmath,amsfonts,amssymb,amsthm}
\usepackage{bm}
\usepackage{graphicx}
\usepackage{booktabs}
\usepackage{cite}
\usepackage[T1]{fontenc}
\usepackage{microtype}
\usepackage[hidelinks]{hyperref}

\newtheorem{proposition}{Proposition}
\theoremstyle{remark}
\newtheorem{remark}{Remark}

\newcommand{\Ybus}{\bm{Y}_{\mathrm{bus}}}
\newcommand{\Vbus}{\bm{V}}
\newcommand{\Sbus}{\bm{S}_{\mathrm{bus}}}
\newcommand{\diag}{\operatorname{diag}}

\newcommand{\midop}{\operatorname{mid}}
\newcommand{\clip}{\operatorname{clip}}
\newcommand{\rank}{\operatorname{rank}}

\begin{document}

\title{Unified Formulation of Voltage Control Constraints for Conventional and Continuation Power Flow via Semismooth Newton and On-Demand Schur Reduction}

\author{Tianshi~Cheng,~\IEEEmembership{Member,~IEEE,}
        and~Venkata~Dinavahi,~\IEEEmembership{Fellow,~IEEE}%
\thanks{This work has been submitted to the IEEE for possible publication. Copyright may be transferred without notice, after which this version may no longer be accessible.}%
\thanks{This work was supported by the Natural Sciences and Engineering Research Council of Canada (NSERC) and Mitacs.}%
\thanks{The authors are with the Department of Electrical and Computer Engineering, University of Alberta, Edmonton, AB T6G 2V4, Canada (e-mail: tcheng1@ualberta.ca; dinavahi@ualberta.ca).}}

\maketitle

\begin{abstract}
Conventional AC power flow has long been challenged by bound constraints such as generator reactive-power limits ($Q_{\lim}$) and remote voltage control (RVC). Existing methods rely on heuristic PV$\leftrightarrow$PQ switching or costly optimization, causing complex implementations, low efficiency, and convergence difficulties in modern renewable-rich grids. This paper observes that $Q_{\lim}$ and RVC are fundamentally voltage control constraints. Formulated as box mixed complementarity problems (Box-MCP) via \texttt{mid} functions, they avoid auxiliary variables and admit single-loop semismooth Newton (SSN) solves without external switching, improving implementation simplicity and convergence. Direct augmentation, however, expands the Jacobian and slows unviolated solves. To address this, an on-demand Schur-complement solver with sparse factor projection is developed around a fixed, shared base Jacobian, preserving original Newton--Raphson matrix size, large-grid efficiency, and batched GPU parallelism. Implemented in RustPower with SuiteSparse KLU up to 9{,}241 buses, the method cuts Newton iterations from 7--15 to 5--9 with Karush--Kuhn--Tucker (KKT) residuals below $1.2\times 10^{-10}$~p.u., speeds up Schur construction by $3.82\times$ over blocked dense solves, and achieves competitive or faster runtimes than C++ lightsim2grid 1.3.1, including $2.4\times$--$2.9\times$ speedups on RVC. Furthermore, appending one continuation port extends the formulation to $Q_{\lim}$- and RVC-bounded continuation power flow (CPF) capable of releasing constraints along the return path; this paper also shows that the reduced Schur complement isolates fold singularity near the nose point and aids bottleneck diagnosis, providing an efficient methodology for steady-state power system analysis.
\end{abstract}


\section*{Nomenclature}
\addcontentsline{toc}{section}{Nomenclature}

\noindent\textbf{A. Sets and Indices}
\begin{IEEEdescription}[\IEEEusemathlabelsep\IEEEsetlabelwidth{$\mathcal{A}_m(\lambda_m)$}]
    \item[$\mathcal{G}_{\mathrm{PV}}$] Local PV generator bus set ($n_{pv} = |\mathcal{G}_{\mathrm{PV}}|$).
    \item[$\mathcal{V},\, \mathcal{I}$] Active (limit-bound) and inactive (in-bounds) subsets of $\mathcal{G}_{\mathrm{PV}}$, with $k_{\mathrm{loc}} = |\mathcal{V}|$.
    \item[$\mathcal{A}_m(\lambda_m)$] Unsaturated generator subset of RVC group $m$.
    \item[$\mathcal{N}(g)$] Adjacent bus set of bus $g$ in the graph of $\Ybus$.
\end{IEEEdescription}

\vspace{2pt}
\noindent\textbf{B. Network, State, and Control Variables}
\begin{IEEEdescription}[\IEEEusemathlabelsep\IEEEsetlabelwidth{$[Q_{\min},Q_{\max}]$}]
    \item[$\Vbus,\, \Ybus$] Complex bus voltage vector ($|V_i|e^{j\theta_i}$) and bus admittance matrix ($G + jB$).
    \item[$x_0 \in \mathbb{R}^{n_0}$] Reduced base state $[\theta;\; |V_{\mathrm{PQ}}|]$, $n_0 = n_{pv} + 2n_{pq}$.
    \item[$v_{\mathcal{V}},\, \lambda_m$] Active Q-limited PV voltage magnitudes ($v_{\mathcal{V}}\in\mathbb{R}^{k_{\mathrm{loc}}}$) and shared scalar of RVC group $m$ ($\lambda\in\mathbb{R}^{k_{\mathrm{rvc}}}$).
    \item[$Q_g,\, V_{\mathrm{set},g}$] Net generator reactive injection and voltage setpoint at bus $g$.
    \item[{$[Q_{\min},Q_{\max}]$}] Reactive-power capability limits of a bus or unit.
    \item[$\alpha_{m,i},\, Q_{G_{m,i}}^0$] Participation factor and reference reactive injection of generator $i$ in RVC group $m$.
    \item[$\gamma_g$] Positive scaling gain ($\gamma_g = \max(1, |B_{gg}|)$).
    \item[$\lambda_{\mathrm{cpf}},\, b_{\mathrm{dir}}$] CPF loading parameter (abbreviated $\lambda$ in plots) and injection increase direction.
\end{IEEEdescription}

\vspace{2pt}
\noindent\textbf{C. Matrices, Factors, and Operators}
\begin{IEEEdescription}[\IEEEusemathlabelsep\IEEEsetlabelwidth{$L_0, U_0, P_r, P_c$}]
    \item[$F_0,\, \Phi$] Base power-mismatch vector and unified Box-MCP control residual vector.
    \item[$J_0,\, J_{\mathrm{aug}}$] Fixed-structure $n_0\times n_0$ base reduced Jacobian and $(n_0+k)\times(n_0+k)$ bordered Newton matrix ($k = k_{\mathrm{loc}} + k_{\mathrm{rvc}}$).
    \item[$U,\, W^\top,\, C$] Low-rank border coupling columns, constraint rows, and $k\times k$ self-coupling block.
    \item[$L_0, U_0, P_r, P_c$] Triangular LU factors and row/column permutations of $J_0$ ($D_R^{-1}P_r J_0 P_c = L_0 U_0$, row scaling $D_R$).
    \item[$X,\, Y$] Half-triangular projection factors ($U_0^\top X = P_c^\top W$, $L_0 Y = D_R^{-1}P_r U$).
    \item[$Z,\, S,\, S_{\mathrm{cpf}}$] Full sensitivity basis ($J_0^{-1}U$), $k\times k$ Schur matrix ($C - X^\top Y$), and $(k+1)\times(k+1)$ CPF Schur matrix.
    \item[$\clip,\,\midop$] Saturation limiter clamping $y$ to $[l,u]$ and median-of-three operator selecting the middle value of $\{a,b,c\}$.
    \item[$M_r^X,\, M_r^Y$] 64-lane active-port bitmask words at row $r$ during wavefront propagation.
\end{IEEEdescription}

\section{Introduction}
\label{sec:introduction}

Enforcing generator reactive-power limits ($Q_{\lim}$) and coordinated remote voltage control (RVC) is central to transmission security assessment and voltage-stability screening, especially in modern grids with high renewable penetration~\cite{popovic1992,kataoka2005,milano2010}. While compiled Newton--Raphson (NR) solvers with direct compressed sparse column (CSC) assembly and in-place refactoring~\cite{davis2010klu,lightsim2grid,cheng2026symbolic} have made unconstrained reduced-Jacobian ($J_0$) iterations very fast, handling voltage control constraints remains a persistent source of implementation complexity, convergence difficulty, and computational overhead. Existing paradigms each sacrifice either compact sparse structure or single-loop convergence:
\begin{enumerate}
    \item \textbf{Outer-loop PV$\leftrightarrow$PQ switching}~\cite{zimmerman2011matpower,thurner2018pandapower} preserves minimal $J_0$ but forces repeated symbolic LU rebuilds and risks one-way clamping traps or active-set cycling.
    \item \textbf{Static full-PQ pre-allocation} (lightsim2grid~\cite{lightsim2grid}) eliminates local-limit symbolic rebuilds by pre-allocating $|V|$ and $Q$ slots for all PV buses, but permanently pays matrix expansion and pivot-growth checks (\texttt{klu\_rgrowth}) even on normal snapshots where zero limits bind ($k = 0$), restarts NR in an outer loop, and resets structure when runtime RVC or continuation ports are added.
    \item \textbf{Global complementarity augmentation}~\cite{pirnia2013,sundaresh2014} embeds local $Q_{\lim}$ via complementarity optimization~\cite{pirnia2013} or augmented NR Jacobians~\cite{sundaresh2014}, but expands the Jacobian across all $n_{\mathrm{pv}}$ generators even at $k = 0$ and omits multi-generator RVC and continuation reuse.
    \item \textbf{External nested RVC loops}~\cite{thurner2018pandapower} wrap full AC solves inside scalar searches, losing quadratic convergence whenever member units saturate sequentially.
\end{enumerate}

This paper observes that local $Q_{\lim}$ and multi-generator RVC are fundamentally voltage control constraints. Formulated as box mixed complementarity (Box-MCP) equations using \texttt{mid} and \texttt{clip} functions without auxiliary variables, they admit single-loop \emph{bordered semismooth Newton (SSN)}~\cite{qisun1993,facchinei2003} solves without external PV$\leftrightarrow$PQ switching, resolving both implementation complexity and convergence reliability. Direct augmentation, however, expands the Jacobian and slows down unviolated operating points, which account for most snapshots in contingency screening and time-series analysis. To preserve the minimal $J_0$ solve when no limits bind while enforcing active controls when needed, this paper introduces \emph{on-demand sparse Schur reduction}~\cite{zhang2005schur,dorfler2013kron,yeung2017amps} around a fixed, shared $J_0$, as illustrated in Fig.~\ref{fig:kron_overview}. At $k = 0$, the solver incurs no extra linear-algebra cost beyond a scalar $Q_g$ bound check; when $k > 0$, only the $k \ll n_0$ active limits, RVC groups, and optional continuation port~\cite{ajjarapu1992,venkatasubramanian1995} are projected onto a compact port matrix $S$. The main contributions of this paper are:
\begin{enumerate}
    \item \textbf{Bordered SSN on minimal $J_0$ (Section~\ref{sec:modeling}):} A Box-MCP border unifying bidirectional local $Q_{\lim}$ and 1-port multi-generator RVC around fixed $J_0$, clarifying the RVC saddle-point polarity that requires outward Schur projection.
    \item \textbf{Sparse Schur properties and $(k+1)\times(k+1)$ CPF reuse (Section~\ref{sec:kron_reduction}):} Factored Schur construction across $J_0$'s LU factors (Proposition~\ref{prop:dual_sparse}) extending symmetric updates~\cite{yeung2017amps} to nonsymmetric SSN active sets, paired with full-PQ step equivalence (Proposition~\ref{prop:full_equiv}), Clique-Closure necessity, anti-chattering scaling, and CPF bordering (Proposition~\ref{prop:spectral_cpf}).
    \item \textbf{Fused sparse Schur implementation (Section~\ref{sec:symbolic_cache}):} Extending the operator-fusion principle of~\cite{cheng2026symbolic} to Schur projection, the implementation combines conjugate-mirror border emission, in-place factor referencing, and bitmask zero-skipping to accumulate the Schur matrix in registers, bypassing intermediate multiple-right-hand-side (multi-RHS) matrices and symbolic reach precomputation~\cite{gilbert1988sparse,amestoy2019mrhs}.
    \item \textbf{Symmetric KLU validation up to 9{,}241 buses (Section~\ref{sec:experiments}):} Warm and cold benchmarks against C++ lightsim2grid (with Python pandapower as a parity reference), confirming zero $J_0$ rebuilds ($3.82\times$ over blocked dense multi-RHS on 9{,}241 buses), $2.4\times$--$2.9\times$ RVC speedups, IEEE~118 limit-induced bifurcation (LIB) tracing, and IEEE~300 bottleneck diagnosis and boundary tracing on the same cached $J_0$.
\end{enumerate}

The remainder of this paper is organized as follows. Section~\ref{sec:modeling} formulates local $Q_{\lim}$ and multi-generator RVC as a unified Box-MCP bordered around fixed $J_0$. Section~\ref{sec:kron_reduction} develops the factored sparse Schur reduction, establishes step equivalence and singularity isolation, and extends the border to $(k+1)\times(k+1)$ CPF. Section~\ref{sec:symbolic_cache} details the conjugate-mirror border emission, bitmask wavefront kernel, and \texttt{SubApp} implementation in RustPower. Section~\ref{sec:experiments} presents numerical accuracy, runtime, ablation, and CPF benchmarks up to 9{,}241 buses, and Section~\ref{sec:conclusion} concludes the paper.

\section{Minimal-$J_0$ Box-Complementarity Modeling and Bordered Newton Formulation}
\label{sec:modeling}

\begin{figure*}[!t]
    \centering
    \includegraphics[width=\textwidth]{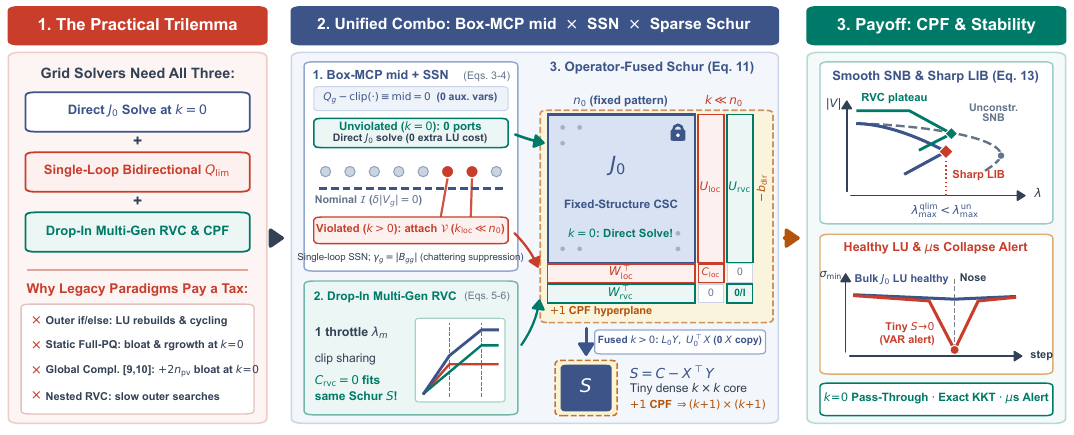}
    \caption{Overview of the reactive-control trilemma and the unified \texttt{mid} + SSN + fused Schur architecture. The \texttt{mid} branches give direct $J_0$ pass-through when unviolated ($k = 0$) and project only active limits ($\mathcal{V}$), RVC groups, and an optional continuation port onto the Schur matrix $S$ without rebuilding $J_0$.}
    \label{fig:kron_overview}
\end{figure*}

Unlike global complementarity models~\cite{pirnia2013,sundaresh2014} that expand the Jacobian across all $n_{\mathrm{pv}}$ generators, this section formulates local $Q_{\lim}$ and multi-generator RVC as piecewise-smooth (\texttt{clip}/\texttt{mid}) equations around a \emph{fixed-pattern reduced Jacobian $J_0$}. Nominal PV buses ($\mathcal{I}$) satisfy $\delta|V_g|=0$ and occupy zero border slots, while only the limit-bound set $\mathcal{V}$ ($k_{\mathrm{loc}} = |\mathcal{V}| \ll n_0$) and configured RVC groups attach as a narrow border (Fig.~\ref{fig:kron_overview}).

\subsection{Pre-Permuted Base Power-Flow Equations}
\label{sec:base_pf}
Consider an $N$-bus AC network with voltages $\Vbus \in \mathbb{C}^N$ ($V_i = |V_i|e^{j\theta_i}$) and admittance $\Ybus = G + jB \in \mathbb{C}^{N\times N}$. Following~\cite{cheng2026symbolic}, buses are pre-permuted into $[\,\mathrm{PQ} \mid \mathrm{PV} \mid \mathrm{slack}\,]$ with counts $n_{pq}$, $n_{pv}$, and $n_{\mathrm{sl}} = N - n_a$ ($n_a = n_{pq} + n_{pv}$); RVC member buses reside in the $\mathrm{PQ}$ partition with reactive output driven by their group scalar. With $\Sbus(\Vbus) = \Vbus \odot \operatorname{conj}(\Ybus \Vbus) = P_{\mathrm{calc}} + j Q_{\mathrm{calc}}$, the base state $x_0 = [\,\theta_{0..n_a-1};\; |V_{0..n_{pq}-1}|\,] \in \mathbb{R}^{n_0}$ ($n_0 = n_{pv} + 2n_{pq}$) and base mismatch are
\begin{equation}
    F_0(x_0, v_{\mathcal{V}}, \lambda) =
    \begin{bmatrix}
        P_{\mathrm{calc},0..n_a-1} - P_{\mathrm{spec},0..n_a-1}\\[2pt]
        Q_{\mathrm{calc},0..n_{pq}-1} - Q_{\mathrm{eff},0..n_{pq}-1}(\lambda)
    \end{bmatrix},
    \label{eq:base_mismatch}
\end{equation}
where $v_{\mathcal{V}} \in \mathbb{R}^{k_{\mathrm{loc}}}$ holds active Q-limited PV voltages, $\lambda \in \mathbb{R}^{k_{\mathrm{rvc}}}$ holds RVC group scalars, and $Q_{\mathrm{eff},i}(\lambda) = Q_{\mathrm{fixed},i} + Q_{G_i}(\lambda)$. The reduced Jacobian $J_0 = \partial F_0/\partial x_0 \in \mathbb{R}^{n_0\times n_0}$ retains a fixed CSC sparsity pattern across all iterations~\cite{cheng2026symbolic}.

\subsection{Local Generator Reactive Limits as a Zero-Expansion Box-MCP}
\label{sec:local_qlim_model}
\begin{figure*}[!t]
    \centering
    \includegraphics[width=0.89\textwidth]{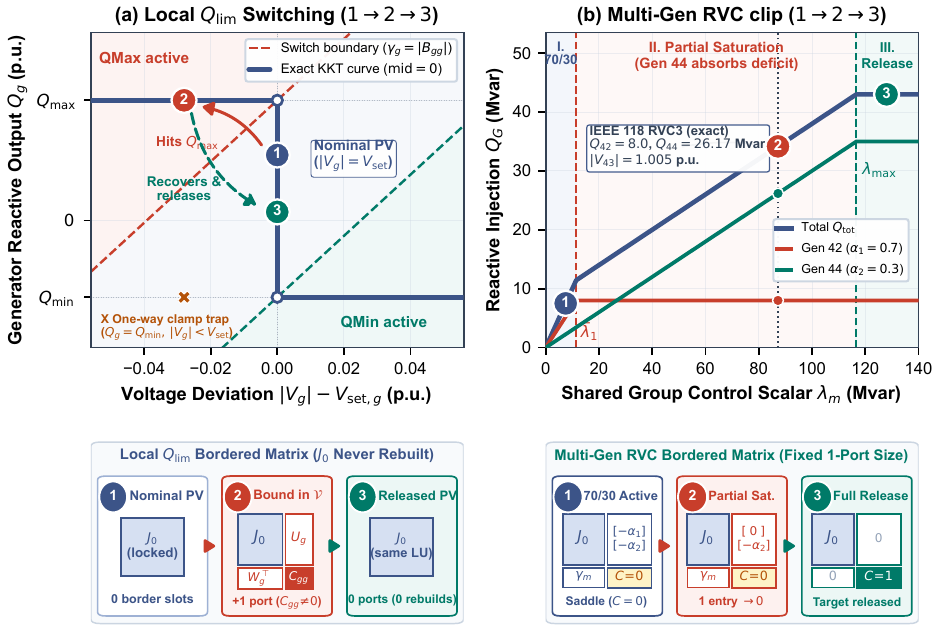}
    \caption{Live switching trajectories ($1 \to 2 \to 3$) synchronized with the bordered Newton matrix around fixed-structure $J_0$. (a)~Local bidirectional Box-MCP ($\gamma_g = |B_{gg}|$): nominal PV occupies zero border slots (State~1), attaches one port $(U_g, W_g^\top, C_{gg})$ at $Q_{\max}$ (State~2), and detaches cleanly upon voltage recovery (State~3). (b)~Coordinated RVC \texttt{clip} sharing on IEEE~118 (Buses~42, 44 $\to$ Bus~43): a fixed 1-port saddle border ($C_{\mathrm{rvc}}=0$, State~1) zeroes $-\alpha_1 \to 0$ under partial saturation (State~2) and flips $C_{\mathrm{rvc}}: 0 \to 1$ upon full release (State~3).}
    \label{fig:boxmcp_rvc}
\end{figure*}

Each local PV bus $g \in \mathcal{G}_{\mathrm{PV}} := \{n_{pq},\ldots,n_a-1\}$ enforces local voltage control using net reactive output $Q_g(x) = Q_{\mathrm{calc},g}(x) - Q_{\mathrm{fixed},g}$ as the control variable: $|V_g|$ is held at $V_{\mathrm{set},g}$ within $[Q_{\min,g}, Q_{\max,g}]$ and released once $Q_g$ saturates, governed by the Karush--Kuhn--Tucker (KKT) conditions~\cite{facchinei2003,pirnia2013,sundaresh2014}
\begin{equation}
\begin{aligned}
    Q_{\min,g} \le Q_g \le Q_{\max,g},\quad
    Q_g < Q_{\max,g} &\implies |V_g| \ge V_{\mathrm{set},g},\\
    Q_g > Q_{\min,g} &\implies |V_g| \le V_{\mathrm{set},g}.
\end{aligned}
    \label{eq:local_kkt}
\end{equation}
Two elementary piecewise-linear operators write~\eqref{eq:local_kkt} as one equation without auxiliary variables~\cite{facchinei2003}: the saturation limiter $\clip(y, l, u) := \min(\max(y, l), u)$, which clamps $y$ to $[l, u]$ (Fig.~\ref{fig:boxmcp_rvc}(b)), and the median-of-three operator $\midop(a,b,c) := \min(\max(a,b), c)$, which returns the middle value of three numbers. With positive gain $\gamma_g > 0$,
\begin{equation}
\begin{aligned}
    \Phi_g(x) &:= Q_g(x) - \clip\!\Bigl(Q_g(x) + \gamma_g\bigl(V_{\mathrm{set},g} - |V_g|\bigr),\\[-2pt]
    &\qquad\qquad\qquad\;\; Q_{\min,g},\; Q_{\max,g}\Bigr)\\
    &\equiv \midop\!\Bigl(\underbrace{Q_g(x) - Q_{\max,g}}_{a_g},\;\; \underbrace{\gamma_g\bigl(|V_g| - V_{\mathrm{set},g}\bigr)}_{b_g},\\[-2pt]
    &\qquad\quad\;\; \underbrace{Q_g(x) - Q_{\min,g}}_{c_g}\Bigr) = 0.
\end{aligned}
    \label{eq:local_mid}
\end{equation}
Because $a_g \le c_g$ everywhere, $\Phi_g(x) = 0$ holds if and only if~\eqref{eq:local_kkt} holds~\cite{facchinei2003}; its zero set traces the classical Z-shaped PV-bus $V$--$Q$ characteristic in Fig.~\ref{fig:boxmcp_rvc}(a). Checking which of $(a_g, b_g, c_g)$ is the middle value splits the $(|V_g|, Q_g)$ plane across the two dashed lines $b_g = a_g$ and $b_g = c_g$ into three smooth regions:
\begin{equation}
    \text{mode}(g) =
    \begin{cases}
        \mathrm{QMax}, & b_g \le a_g \;\;(\Phi_g = a_g),\\
        \mathrm{QMin}, & b_g \ge c_g \;\;(\Phi_g = c_g),\\
        \mathrm{PV},   & a_g < b_g < c_g \;\;(\Phi_g = b_g).
    \end{cases}
    \label{eq:local_modes}
\end{equation}
Because $\Phi_g$ is piecewise smooth, each SSN iteration simply picks the active branch in~\eqref{eq:local_modes} and takes its ordinary derivative~\cite{qisun1993,facchinei2003}, driving the bus automatically through States~$1 \to 2 \to 3$ in Fig.~\ref{fig:boxmcp_rvc}(a) without external PV$\leftrightarrow$PQ switching:
\begin{itemize}
    \item \emph{State~1 (Nominal PV, $g \in \mathcal{I}$):} $b_g$ is the middle value, so $\nabla_x \Phi_g = \gamma_g e_{|V_g|}^\top$ forces $\delta|V_g| = 0$; the bus stays at $V_{\mathrm{set},g}$ and needs zero border slots in $J_0$.
    \item \emph{State~2 (Limit-bound, $g \in \mathcal{V}$):} crossing the upper dashed line ($b_g \le a_g$) makes $a_g$ active, so SSN switches the row to $\nabla_x \Phi_g = \partial Q_g/\partial x$ and attaches one border port to release $|V_g|$.
    \item \emph{State~3 (Released PV):} if voltage recovers across the dashed line back into $a_g < b_g < c_g$, $b_g$ becomes active again, detaching the port and restoring PV control cleanly instead of locking in a one-way clamp trap (orange cross in Fig.~\ref{fig:boxmcp_rvc}(a)).
\end{itemize}

\subsection{Coordinated Multi-Generator Remote Voltage Control}
\label{sec:rvc_model}
When $p_m \ge 1$ generators $\{g_{m,1},\ldots,g_{m,p_m}\} \subset \{0,\ldots,n_{pq}-1\}$ regulate remote bus $r_m$ to $V_{\mathrm{set},m}$ with participation factors $\alpha_{m,i} > 0$ ($\sum_i \alpha_{m,i} = 1$)~\cite{popovic1992}, one shared scalar $\lambda_m \in \mathbb{R}$ drives all $p_m$ units from interior references $Q_{G_{m,i}}^0 \in (Q_{\min,m,i}, Q_{\max,m,i})$ through individual saturation limiters:
\begin{equation}
    Q_{G_{m,i}}(\lambda_m) := \clip\!\bigl(Q_{G_{m,i}}^0 + \alpha_{m,i}\lambda_m,\;\; Q_{\min,m,i},\;\; Q_{\max,m,i}\bigr).
    \label{eq:rvc_clip}
\end{equation}
With unit breakpoints $\underline{\lambda}_{m,i} = (Q_{\min,m,i} - Q_{G_{m,i}}^0)/\alpha_{m,i} \le 0 \le (Q_{\max,m,i} - Q_{G_{m,i}}^0)/\alpha_{m,i} = \bar{\lambda}_{m,i}$ and group bounds $\lambda_{\min,m} = \min_i \underline{\lambda}_{m,i}$, $\lambda_{\max,m} = \max_i \bar{\lambda}_{m,i}$, each interval $[\underline{\lambda}_{m,i}, \bar{\lambda}_{m,i}]$ contains $0$, so their union $[\lambda_{\min,m}, \lambda_{\max,m}]$ is a single gap-free interval on which the unsaturated subset $\mathcal{A}_m(\lambda_m) := \{i : \underline{\lambda}_{m,i} < \lambda_m < \bar{\lambda}_{m,i}\}$ is nonempty for all $\lambda_m \in (\lambda_{\min,m}, \lambda_{\max,m})$ and maintains exact proportional sharing $\Delta Q_{G_{m,i}}/\Delta Q_{G_{m,j}} = \alpha_{m,i}/\alpha_{m,j}$. Remote regulation with full-saturation release follows the same $\midop$ form as~\eqref{eq:local_mid}:
\begin{equation}
\begin{aligned}
    \Phi_m^{\mathrm{rvc}}(x, \lambda_m) := {}&\midop\Bigl(\lambda_m - \lambda_{\max,m},\\[-2pt]
    &\gamma_m\bigl(|V_{r_m}| - V_{\mathrm{set},m}\bigr),\; \lambda_m - \lambda_{\min,m}\Bigr) = 0.
\end{aligned}
    \label{eq:rvc_mid}
\end{equation}
Because~\eqref{eq:rvc_clip} is piecewise linear in $\lambda_m$, its slope is simply $\partial Q_{G_{m,i}}/\partial \lambda_m = \alpha_{m,i}$ on the ramp ($i \in \mathcal{A}_m(\lambda_m)$) and $0$ on the flat saturated plateau. Following States~$1 \to 2 \to 3$ in Fig.~\ref{fig:boxmcp_rvc}(b), SSN shares reactive demand proportionally while both units are unsaturated (State~1), zeroes a saturated unit's slope so remaining units automatically absorb the deficit (State~2), and switches~\eqref{eq:rvc_mid} to its bound branch ($\lambda_m = \lambda_{\max,m}$, flipping $C_{\mathrm{rvc}}: 0 \to 1$) to release $|V_{r_m}|$ upon full saturation (State~3). At the endpoints $\lambda_m \in \{\lambda_{\min,m}, \lambda_{\max,m}\}$, if the group re-enters regulation from full saturation, the inward one-sided Clarke derivative replaces $\mathcal{A}_m(\lambda_m)$ by $\{i : \underline{\lambda}_{m,i} \le \lambda_m \le \bar{\lambda}_{m,i}\}$, preventing a spurious zero column in~\eqref{eq:u_rvc}.

\subsection{Bordered Generalized Newton Matrix and Structural Distinction}
\label{sec:bordered_matrix}
With $k = k_{\mathrm{loc}} + k_{\mathrm{rvc}}$, $\delta_u = [\delta v_{\mathcal{V}};\, \delta\lambda] \in \mathbb{R}^k$, and $\Phi = [\Phi_{\mathrm{loc}};\, \Phi_{\mathrm{rvc}}] \in \mathbb{R}^k$, the SSN step solves the $(n_0+k)\times(n_0+k)$ bordered system
\begin{equation}
\begin{aligned}
    \underbrace{\begin{bmatrix}
        J_0 & U_{\mathrm{loc}} & U_{\mathrm{rvc}}\\
        W_{\mathrm{loc}}^\top & C_{\mathrm{loc}} & 0\\
        W_{\mathrm{rvc}}^\top & C_{\mathrm{rvc,loc}} & C_{\mathrm{rvc}}
    \end{bmatrix}}_{J_{\mathrm{aug}}}
    \begin{bmatrix}
        \delta_0\\ \delta v_{\mathcal{V}}\\ \delta\lambda
    \end{bmatrix}
    &= -\begin{bmatrix} F_0\\ \Phi_{\mathrm{loc}}\\ \Phi_{\mathrm{rvc}} \end{bmatrix}\\[2pt]
    \iff
    \begin{bmatrix} J_0 & U\\ W^\top & C \end{bmatrix}
    \begin{bmatrix} \delta_0\\ \delta_u \end{bmatrix}
    &= -\begin{bmatrix} F_0\\ \Phi \end{bmatrix},
\end{aligned}
    \label{eq:bordered_compact}
\end{equation}
whose sparse border blocks follow directly from~\eqref{eq:base_mismatch}--\eqref{eq:rvc_mid}:
\begin{enumerate}
    \item \textbf{Local column $U_{\mathrm{loc}}$ and row $(W_{\mathrm{loc}}^\top, C_{\mathrm{loc}})$ ($g = \mathcal{V}_j$):} $(U_{\mathrm{loc}})_{:,j} = \partial F_0/\partial |V_g|$ and $(W_{\mathrm{loc}}^\top, C_{\mathrm{loc}})_{j,:} = \partial Q_g/\partial(x_0, v_{\mathcal{V}})$, supported solely on $g$ and its $\Ybus$ neighbors $\mathcal{N}(g)$.
    \item \textbf{RVC column $U_{\mathrm{rvc}}$ and row $(W_{\mathrm{rvc}}^\top, C_{\mathrm{rvc,loc}}, C_{\mathrm{rvc}})$:}
    \begin{equation}
        (U_{\mathrm{rvc}})_{:,m} = -\!\sum_{i \in \mathcal{A}_m(\lambda_m)} \!\alpha_{m,i}\, e_{n_a + g_{m,i}},
        \label{eq:u_rvc}
    \end{equation}
    which vanishes when $\mathcal{A}_m(\lambda_m) = \emptyset$. While regulating ($\lambda_{\min,m} < \lambda_m < \lambda_{\max,m}$), $(C_{\mathrm{rvc}})_{m,m} = 0$ and row $m$ has a single $\gamma_m$ entry at $n_a + r_m$ (in $W_{\mathrm{rvc}}^\top$ if $r_m < n_{pq}$, or $(C_{\mathrm{rvc,loc}})_{m,l}$ if $r_m = \mathcal{V}_l$; if $r_m \in \mathcal{I}$, row $m$ vanishes, exposing the setpoint conflict). When saturated, row $m$ becomes $e_{k_{\mathrm{loc}}+m}^\top$ ($(C_{\mathrm{rvc}})_{m,m} = 1$).
\end{enumerate}

Table~\ref{tab:rvc_vs_qlim} contrasts local $Q_{\lim}$ and coordinated RVC within~\eqref{eq:bordered_compact}. Whenever $k = 0$ ($\mathcal{V} = \emptyset$, no active RVC), $(U, W^\top, C)$ vanish and~\eqref{eq:bordered_compact} reduces identically to the unconstrained step $J_0\delta_0 = -F_0$, preserving nominal $J_0$ speed without pre-allocation tax; when $k > 0$, outward Schur projection replaces $(C_{\mathrm{rvc}})_{mm} = 0$ with the closed-network sensitivity $\gamma_m(\mathrm{d}|V_{r_m}|/\mathrm{d}\lambda_m)|_{F_0=0}$.

\begin{table*}[!t]
\centering
\caption{Structural and Algebraic Distinction Between Local Generator $Q_{\lim}$ and Coordinated Multi-Generator RVC in~\eqref{eq:bordered_compact}}
\label{tab:rvc_vs_qlim}
\renewcommand{\arraystretch}{1.08}
\begin{tabular}{@{}p{2.7cm}p{7.0cm}p{7.2cm}@{}}
\toprule
\textbf{Property} & \textbf{Local Generator $Q_{\lim}$ ($g \in \mathcal{G}_{\mathrm{PV}}$)} & \textbf{Coordinated Multi-Generator RVC (Group $m$)} \\
\midrule
Active-set polarity & Nominal PV is \emph{inactive} (0 border slots); saturation \emph{attaches} a port releasing $|V_g|$. & Nominal regulation is \emph{active} (shared $\lambda_m$); full saturation \emph{releases} $|V_{r_m}|$. \\
Self-block $C$ & Nonsingular diagonal $(C_{\mathrm{loc}})_{jj} = \partial Q_g/\partial|V_g| \approx |B_{gg}| \neq 0$. & Saddle-point $(C_{\mathrm{rvc}})_{mm} = 0$ while regulating ($J_0 - U C^{-1} W^\top$ undefined). \\
Spatial coupling & Collocated at bus $g$; $U_{\mathrm{loc}}$ and $W_{\mathrm{loc}}^\top$ share $\mathcal{N}(g)$ in $\Ybus$. & Non-collocated ($p_m$ units $\{g_{m,i}\}$ drive remote $r_m$); $U_{\mathrm{rvc}}$ and $W_{\mathrm{rvc}}^\top$ disjoint. \\
Saturation & Single-stage switching ($\mathrm{PV} \leftrightarrow \mathrm{QMax}/\mathrm{QMin}$). & Multi-stage: units saturate sequentially, zeroing $\alpha_{m,i}$ in~\eqref{eq:u_rvc}. \\
\bottomrule
\end{tabular}
\end{table*}

\section{Sparse Schur Construction and Reduced-Space Stability Properties}
\label{sec:kron_reduction}

Eliminating an interior domain onto a small port interface via Schur complements~\cite{zhang2005schur,golub2013matrix} is the Jacobian counterpart to network Kron reduction~\cite{dorfler2013kron}. Rather than forming a dense reduction across all generators, the proposed method keeps bulk $J_0 \in \mathbb{R}^{n_0\times n_0}$ sparse and invariant while performing \emph{sparse Schur construction} onto only the $k \ll n_0$ active ports.

\subsection{Outward Block Elimination and Clique-Closure Necessity}
\label{sec:block_elim}
For nonsingular $J_0$, left-multiplying~\eqref{eq:bordered_compact} by $L_{\mathrm{blk}} = \bigl[\begin{smallmatrix} I_{n_0} & 0 \\ -W^\top J_0^{-1} & I_k \end{smallmatrix}\bigr]$ ($\det L_{\mathrm{blk}} = 1$) yields the block-triangular reduction
\begin{equation}
    L_{\mathrm{blk}}\, J_{\mathrm{aug}} =
    \begin{bmatrix} J_0 & U\\ 0 & S \end{bmatrix},
    \qquad
    S := C - W^\top J_0^{-1} U \in \mathbb{R}^{k \times k},
    \label{eq:schur_def}
\end{equation}
decoupling~\eqref{eq:bordered_compact} into $J_0[\delta_0^{(0)}, Z] = [-F_0, U]$, $S\delta_u = -\Phi - W^\top \delta_0^{(0)}$, and $\delta_0 = \delta_0^{(0)} - Z\delta_u$. Physically, $S_{ij} = \mathrm{d}Q_{\mathcal{V}_i}/\mathrm{d}|V_{\mathcal{V}_j}||_{F_0=0}$ gives closed-network reactive stiffness between saturated PV ports, while $S_{k_{\mathrm{loc}}+m,\,k_{\mathrm{loc}}+l} = \gamma_m \mathrm{d}|V_{r_m}|/\mathrm{d}\lambda_l|_{F_0=0}$ (since $\mathrm{d}x_0/\mathrm{d}\lambda_l = -J_0^{-1}(U_{\mathrm{rvc}})_{:,l}$) gives remote voltage sensitivity even when $(C_{\mathrm{rvc}})_{mm} = 0$.

\begin{remark}[Degree and Clique-Closure Criterion for outward elimination]
\label{rem:clique_closure}
Even for local $Q_{\lim}$ where $c_g = (C_{\mathrm{loc}})_{jj} \neq 0$, folding a degree-$d_g$ bus $g$ ($d_g = |\mathcal{N}(g)|$) inward into $J_0$ via $\widetilde{J}_0 = J_0 - c_g^{-1} U_g W_g^\top$ couples all pairs in $(\mathcal{N}(g)\cup\{g\})^2$, injecting $\binom{d_g}{2} - |E(\mathcal{N}(g))|$ off-diagonal entries unless $\mathcal{N}(g)$ forms a clique in $\Ybus$ ($d_g \le 1$ step-up leaves or $d_g \ge 2$ closed triangles). Because high-voltage equivalents absorb step-up transformers into meshed substations ($\bar{d}_g = 3.9$--$5.6$ across saturated PV buses, leaving $88\%$--$100\%$ non-clique in Fig.~\ref{fig:cache_bloat}(b)), outward Schur projection onto $S$ is the general sparsity-preserving direction.
\end{remark}

\subsection{Factored Sparse Schur Construction}
\label{sec:dual_sparse_kron}
Materializing $Z = J_0^{-1}U \in \mathbb{R}^{n_0\times k}$ via dense multi-RHS solves produces a typically dense matrix occupying $8n_0 k$ bytes ($27.6$~MB on PEGASE~9241 at $k=203$) and requires $k$ backward solves of $U_0$. Splitting the projection across $J_0$'s LU factors restricts propagation to the half-triangular spaces $(X, Y)$, where $95\%$--$97.4\%$ of entries remain zero:

\begin{proposition}[Factored sparse Schur identity]
\label{prop:dual_sparse}
Let $D_R^{-1} P_r J_0 P_c = L_0 U_0$ be the scaled LU factorization of nonsingular $J_0$, with permutations $P_r, P_c$, row scaling $D_R \succ 0$, lower-triangular $L_0$, and upper-triangular $U_0$. Defining $Y, X \in \mathbb{R}^{n_0\times k}$ by
\begin{equation}
    L_0 Y = D_R^{-1} P_r U, \qquad U_0^\top X = P_c^\top W,
    \label{eq:dual_sparse_xy}
\end{equation}
the Schur matrix~\eqref{eq:schur_def} and base state recovery satisfy
\begin{equation}
\begin{aligned}
    S &= C - X^\top Y = C - \sum_{r=0}^{n_0-1} X_{r,:}^\top Y_{r,:},\\
    \delta_0 &= \delta_0^{(0)} - P_c U_0^{-1}\bigl(Y\,\delta_u\bigr),
\end{aligned}
    \label{eq:factored_kron}
\end{equation}
requiring zero $U_0$ solves during $S$ formation and a \emph{single} backward solve $U_0\hat{z} = Y\delta_u$ for state recovery.
\end{proposition}

\begin{proof}
Since $J_0^{-1} = P_c U_0^{-1} L_0^{-1} D_R^{-1} P_r$, substituting~\eqref{eq:dual_sparse_xy} yields $W^\top J_0^{-1} U = (U_0^{-\top} P_c^\top W)^\top (L_0^{-1} D_R^{-1} P_r U) = X^\top Y$ and $J_0^{-1} U \delta_u = P_c U_0^{-1}(Y\delta_u)$.
\end{proof}

Whereas Yeung \emph{et al.}~\cite{yeung2017amps} (AMPS) used bordered Schur complements with precomputed symbolic reach for symmetric $N{-}k$ admittance updates, \eqref{eq:factored_kron} applies to nonsymmetric Jacobians ($J_0 \neq J_0^\top$, $U \neq W$), changing active sets $\mathcal{V}^{(\nu)}$, and saddle-point RVC ($C_{\mathrm{rvc}}=0$). Moreover, under a border-last elimination ordering, monolithic LU incurs the same active-port fill-in at the root of its elimination tree; externalizing this $k\times k$ front $S$ (factored with partial pivoting) decouples mode switching from bulk $(L_0, U_0)$, keeping the core solver modular, read-only, and free of row overwrites or pivot-growth rescans (\texttt{klu\_rgrowth}).

\subsection{Full-PQ Step Equivalence and Anti-Chattering Scaling}
\label{sec:gamma_scaling}
\begin{proposition}[Full-PQ step equivalence]
\label{prop:full_equiv}
Assume $J_0$ is nonsingular. When any releasing bus $g \in \mathcal{I}$ is snapped to $|V_g| = V_{\mathrm{set},g}$ prior to evaluating $(F_0, J_0)$, the step $(\delta_0, \delta_u)$ from~\eqref{eq:factored_kron} with $\delta v_{\mathcal{I}} = \bm{0}$ is algebraically identical to the $(2n_a + k_{\mathrm{rvc}})$-dimensional full-PQ generalized Newton step while keeping $J_0$'s CSC pattern invariant.
\end{proposition}

\begin{proof}
In the full-PQ ordering $[x_0;\, u;\, v_{\mathcal{I}}]$, the inactive block row is $\Gamma_{\mathcal{I}}\delta v_{\mathcal{I}} = -\Gamma_{\mathcal{I}}(v_{\mathcal{I}} - v_{\mathcal{I}}^{\mathrm{set}})$ with $\Gamma_{\mathcal{I}} = \diag(\gamma_g)_{g\in\mathcal{I}} \succ 0$:
\begin{equation}
    \begin{bmatrix}
        J_0 & U & U_{\mathcal{I}}\\
        W^\top & C & C_{u,\mathcal{I}}\\
        0 & 0 & \Gamma_{\mathcal{I}}
    \end{bmatrix}
    \begin{bmatrix} \delta_0\\ \delta_u\\ \delta v_{\mathcal{I}} \end{bmatrix}
    = -\begin{bmatrix} F_0\\ \Phi\\ \Gamma_{\mathcal{I}}(v_{\mathcal{I}} - v_{\mathcal{I}}^{\mathrm{set}}) \end{bmatrix}.
    \label{eq:full_3x3}
\end{equation}
With $v_{\mathcal{I}} = v_{\mathcal{I}}^{\mathrm{set}}$, $\delta v_{\mathcal{I}} = \bm{0}$ eliminates the third block column identically, reducing~\eqref{eq:full_3x3} to~\eqref{eq:bordered_compact} (even without snapping, substituting $\delta v_{\mathcal{I}} = v_{\mathcal{I}}^{\mathrm{set}} - v_{\mathcal{I}}$ merely shifts $(-F_0, -\Phi)$ by $-(U_{\mathcal{I}}, C_{u,\mathcal{I}})\delta v_{\mathcal{I}}$ without altering $J_0$).
\end{proof}

Because $F_0$ is smooth and $\Phi$ is strongly semismooth (piecewise $C^2$), BD-regularity yields local quadratic SSN convergence~\cite{qisun1993,facchinei2003}. Furthermore, rearranging the $\mathrm{QMax} \to \mathrm{PV}$ release test $a_g < b_g$ in~\eqref{eq:local_modes} at $|V_g| < V_{\mathrm{set},g}$ gives $Q_g(|V_g|) + \gamma_g(V_{\mathrm{set},g} - |V_g|) < Q_{\max,g}$. At fixed neighboring states in inductive grids ($B_{gg} < 0$, $|V_g|\approx 1$~p.u.), $\partial Q_g/\partial|V_g| \approx |B_{gg}|$; choosing $\gamma_g = \max(1, |B_{gg}|)$ therefore suppresses release-and-reclamp chattering at both $Q_{\max,g}$ and $Q_{\min,g}$.

\subsection{Singularity Isolation and Unified $(k+1)\times(k+1)$ CPF}
\label{sec:lib_cpf}
Scaling injections along $b_{\mathrm{dir}} \in \mathbb{R}^{n_0}$ via $F_0^{\mathrm{base}} - \lambda_{\mathrm{cpf}} b_{\mathrm{dir}}$ ($c_\lambda = \partial\Phi/\partial\lambda_{\mathrm{cpf}}$) with pseudo-arclength constraint $g_{\mathrm{arc}}(x_0, u, \lambda_{\mathrm{cpf}}) = 0$ (tangent $[t_{x_0}^\top, t_u^\top, t_\lambda]$) appends one border column $-b_{\mathrm{dir}}$ and one row to~\eqref{eq:bordered_compact}, yielding the $(k+1)\times(k+1)$ CPF Schur matrix~\cite{ajjarapu1992}:
\begin{equation}
    S_{\mathrm{cpf}} =
    \begin{bmatrix}
        S & c_\lambda + W^\top J_0^{-1} b_{\mathrm{dir}}\\
        t_u^\top - t_{x_0}^\top Z & t_\lambda + t_{x_0}^\top J_0^{-1} b_{\mathrm{dir}}
    \end{bmatrix}.
    \label{eq:scpf}
\end{equation}

\begin{proposition}[Port singularity isolation and CPF bordering]
\label{prop:spectral_cpf}
Whenever bulk $J_0$ is nonsingular ($\det J_0 \neq 0$), $\det J_{\mathrm{aug}} = \det J_0 \det S$ concentrates bifurcation singularity inside the $k\times k$ Schur matrix $S$; at a simple saddle-node bifurcation (SNB) fold ($\rank S = k-1$ with left/right null vectors $\psi, \phi$), rank-1 bordering~\eqref{eq:scpf} ($\det J_{\mathrm{cpf}} = \det J_0 \det S_{\mathrm{cpf}}$) is nonsingular whenever $\psi^\top(c_\lambda + W^\top J_0^{-1} b_{\mathrm{dir}}) \neq 0$ and $(t_u^\top - t_{x_0}^\top Z)\phi \neq 0$~\cite{zhang2005schur,ajjarapu1992,govaerts2000}.
\end{proposition}

Because Q-saturation removes local voltage support, Q-limited collapse typically precedes the unconstrained nose ($\lambda_{\max}^{\mathrm{qlim}} < \lambda_{\max}^{\mathrm{un}}$)~\cite{kataoka2005,venkatasubramanian1995}; in this typical regime, bulk $J_0$ remains well-conditioned while $S$ captures the loss of reactive stiffness, though this is an empirical property rather than a general guarantee (atypical $\lambda_{\max}^{\mathrm{qlim}} \gtrsim \lambda_{\max}^{\mathrm{un}}$ cases revert to Keller bordering~\cite{govaerts2000}). Practically, monitoring $S$'s conditioning or collapse mode along a stalled solve helps pinpoint bottleneck generators for limit relaxation or trigger an early switch into $(k+1)\times(k+1)$ CPF while reusing fixed $J_0$ throughout.

\section{Operator-Fused Sparse Schur Implementation and SubApp Architecture}
\label{sec:symbolic_cache}

\begin{figure*}[!t]
    \centering
    \includegraphics[width=0.90\textwidth]{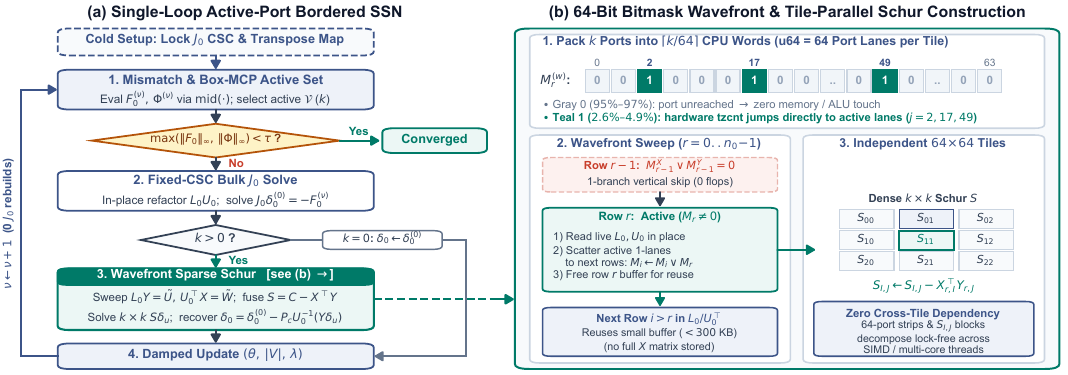}
    \caption{Bordered SSN and fused sparse Schur construction. (a)~Single-loop SSN iteration around fixed $J_0$, linking Step~3 directly to the fused Schur kernel. (b)~Bitmask zero-skipping and lock-free $64\times 64$ tile accumulation ($S_{I,J} \leftarrow S_{I,J} - X_{r,I}^\top Y_{r,J}$), eliminating intermediate factor matrices.}
    \label{fig:ssn_flowchart}
\end{figure*}

\subsection{Conjugate-Mirror Sparse Border Emission}
\label{sec:mirror_assembly}
In the pre-permuted $[\mathrm{PQ}\mid\mathrm{PV}\mid\mathrm{slack}]$ CSC layout of~\cite{cheng2026symbolic}, each column $g$ of $\Ybus$ is split by precomputed pointers $(q_g, a_g, d_g)$ into PQ $[0, q_g)$, PV $[q_g, a_g)$, and slack $[a_g, d_g)$ slices, paired with a transpose slot map $m(p)$ giving $Y_{g,i} = \Ybus[m(p)]$. For each active bus $g \in \mathcal{V}$, sweeping $[0, q_g)$ and $[q_g, a_g)$ emits sparse tuples of $(U_{\mathrm{loc}})_{:,j}$ directly in $O(|\mathcal{N}(g)|)$ work without zeroing $n_0$-vectors, while $C(\Ybus, \Vbus)^\top = C(\Ybus^\top, \operatorname{conj}(\Vbus))$ and $m(p)$ emit row $j$ of $(W_{\mathrm{loc}}^\top, C_{\mathrm{loc}})$ in the same pass without coordinate sorting. Local $Q_{\lim}$ activates once $\|F_0\|_\infty < \tau_{\mathrm{act}}$ (default $0.5$~p.u.).

\subsection{Bitmask Wavefront and Tile-Parallel Schur Construction}
\label{sec:cache_crossover}
Exploiting vertical and horizontal multi-RHS sparsity normally requires Gilbert--Peierls depth-first reachability~\cite{gilbert1988sparse} or column-reordering trees~\cite{amestoy2019mrhs}, adding overhead when $\mathcal{V}^{(\nu)}$ changes every SSN step. Instead, the solver fuses reach discovery, numerical propagation, and outer-product accumulation in~\eqref{eq:factored_kron} into one wavefront operator (Fig.~\ref{fig:ssn_flowchart}(b)):
\begin{enumerate}
    \item \textbf{Cold-cached $U_0^\top$ offsets:} Because \texttt{refactor} preserves $(P_r, P_c)$ and factor sparsity, compressed-column offsets of lower-triangular $U_0^\top$ into the factor buffer are cached once per cold \texttt{factor}, reading live $(L_0, U_0, D_R)$ values in place with zero extraction copies.
    \item \textbf{Bitmask wavefront and lock-free tiling:} Grouping $k$ ports into $\lceil k/64\rceil$ CPU words (\texttt{u64}, one bit per port lane) inside a $<300$~KB L2 slot pool skips unreached rows ($M_r^X = M_r^Y = 0$) in one branch and uses hardware \texttt{tzcnt} bit scans to touch only nonzero port lanes ($2.6\%$--$4.9\%$ fill), fusing $S \mathrel{-}= X_{r,:}^\top Y_{r,:}$ in registers (\emph{zero $X$ storage}). Crucially, port columns in~\eqref{eq:dual_sparse_xy} have zero cross-lane data dependency: 64-port strips and $64\times 64$ Schur tiles $S_{I,J} = C_{I,J} - X_{:,I}^\top Y_{:,J}$ decompose independently without synchronization across vector lanes or multi-core threads.
\end{enumerate}

\subsection{SubApp Isolation and Open-Source Implementation}
\label{sec:subapp_arch}
In RustPower~\cite{rustpower}, the $Q_{\lim}$, RVC, and $(k+1)\times(k+1)$ CPF engine runs inside an isolated \texttt{SubApp} that borrows the source \texttt{World} read-only once and owns all symbolic caches and wavefront buffers, imposing zero overhead on unconstrained solves; full kernel and damping source code is openly available in~\cite{rustpower}.

\section{Numerical Experiments}
\label{sec:experiments}

\begin{table}[!t]
\centering
\caption{Evaluated Solver Configurations and Linear Backends}
\label{tab:solvers}
\footnotesize
\setlength{\tabcolsep}{3.0pt}
\renewcommand{\arraystretch}{1.10}
\begin{tabular}{@{}p{2.12cm}p{3.33cm}p{2.80cm}@{}}
\toprule
\textbf{Solver} & \textbf{Control \& Jacobian} & \textbf{Linear Solver} \\
\midrule
RustPower~\cite{rustpower} & Active-port SSN + fused Schur & 64-bit KLU (\texttt{klu\_l\_*}), W/C \\
lightsim2grid 1.3.1~\cite{lightsim2grid} & Static Full-PQ + \texttt{ReactiveLimits} & C++ KLU (\texttt{NROuter\_KLU}), W/C \\
RustPower Legacy & Outer PV$\to$PQ rebuild & 64-bit KLU, cold \\
pandapower 3.5.4~\cite{thurner2018pandapower} & Outer PV$\to$PQ + \texttt{BinarySearchControl} & SciPy SuperLU (Python ref.) \\
\bottomrule
\end{tabular}
\par\vspace{2pt}
\begin{minipage}{\columnwidth}
\scriptsize W/C: symmetric warm median (cached symbolic analysis) and cold median (\texttt{reset()} / \texttt{tell\_solver\_need\_reset()}). Short labels ls2g and pp denote lightsim2grid and pandapower in Tables~\ref{tab:suite1_qlim}--\ref{tab:suite2_rvc}.
\end{minipage}
\end{table}

\begin{table*}[!t]
\centering
\caption{Suite 1: Local Generator $Q_{\lim}$ Symmetric Warm/Cold Speed, KKT Fidelity, and Parity ($\tau = 10^{-8}$~p.u., SuiteSparse KLU)}
\label{tab:suite1_qlim}
\setlength{\tabcolsep}{4.4pt}
\renewcommand{\arraystretch}{1.08}
\begin{tabular}{@{}lrrrrrrrrr@{}}
\toprule
\textbf{Benchmark Grid} & \textbf{Active} & \textbf{RustPower} & \textbf{NR} & \textbf{ls2g 1.3.1 C++} & \textbf{ls2g} & \textbf{Legacy} & \textbf{pandapower} & \textbf{$\max|\Delta V|$} & \textbf{KKT Viol.} \\
 & $Q^+ / Q^-$ & \textbf{W / C (ms)} & \textbf{It.} & \textbf{W / C (ms)} & \textbf{It.} & \textbf{(ms)} & \textbf{SciPy (ms)} & \textbf{vs.\ pp (p.u.)} & \textbf{(O/l/p)} \\
\midrule
IEEE 118       & $1 / 5$     & $\mathbf{0.18} / \mathbf{0.29}$  & 5 & $0.32 / 0.62$     &  7 &   3.76 &  59.1 & $9.99\times 10^{-16}$ & \textbf{0/0/0} \\
Illinois 200   & $37 / 0$    & $1.03 / \mathbf{1.15}$           & 9 & $\mathbf{0.92} / 1.27$ & 14 &   5.81 &  94.3 & $1.66\times 10^{-9}$  & \textbf{0/0/0} \\
PEGASE 1354    & $25 / 0$    & $\mathbf{2.81} / \mathbf{4.20}$  & 6 & $5.63 / 7.74$     & 10 &  22.24 & 124.1 & $2.33\times 10^{-11}$ & \textbf{0/0/0} \\
PEGASE 2869    & $72 / 0$    & $\mathbf{10.93} / \mathbf{14.04}$ & 7 & $19.49 / 25.47$   & 13 &  59.68 & 244.0 & $7.02\times 10^{-14}$ & \textbf{0/0/0} \\
PEGASE 9241    & $196 / 7$   & $\mathbf{76.58} / \mathbf{102.65}$ & 8 & $103.63 / 122.67$ & 15 & 253.04 & 965.6 & $7.86\times 10^{-5}\dagger$ & \textbf{0/0/1} \\
\midrule
IEEE 300       & \multicolumn{9}{l}{\emph{Connected Q-limited operating branch folds before $\lambda = 1.0$; CPF traces $\lambda_{\max} = 0.98139 < 1.0$ (Section~\ref{sec:exp_cpf}).}} \\
\bottomrule
\end{tabular}
\par\vspace{2pt}
\begin{minipage}{\textwidth}
\scriptsize $\dagger$One-way pandapower and Legacy QLim over-clamp Bus~6752 at $Q_{\min}=-14.02$~Mvar ($|V| < V_{\mathrm{set}}$); both RustPower (O) and lightsim2grid 1.3.1 (l) release Bus~6752 cleanly (0 violations).
\end{minipage}
\end{table*}

\begin{figure*}[!t]
    \centering
    \includegraphics[width=0.92\textwidth]{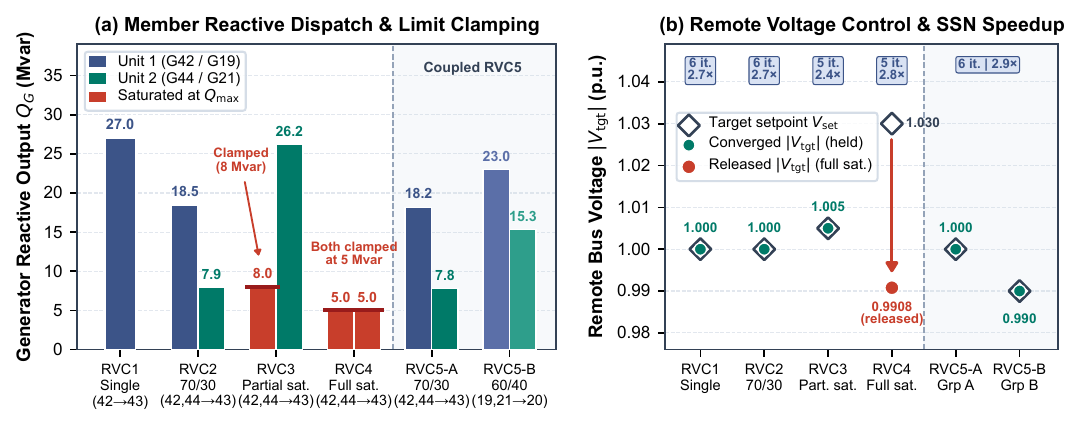}
    \caption{Physical operating points across IEEE~118 RVC1--RVC5 scenarios (simultaneous with local $Q_{\lim}$, $\tau = 10^{-10}$~p.u.). (a)~Converged reactive dispatch showing exact 70/30 and 60/40 sharing (RVC2, RVC5), partial saturation redistribution (RVC3: Bus~42 clamped at $8$~Mvar, Bus~44 absorbing $26.2$~Mvar), and full clamping (RVC4: both at $5$~Mvar). (b)~Remote target voltage $|V_{\mathrm{tgt}}|$ vs.\ $V_{\mathrm{set}}$, showing exact regulation in RVC1--RVC3 and RVC5, clean voltage release in RVC4 ($0.9908$~p.u.), and single-loop SSN iterations and warm speedups over C++ lightsim2grid 1.3.1.}
    \label{fig:benchmarks}
\end{figure*}

\subsection{Experimental Setup and Fairness Protocol}
\label{sec:exp_setup}
All benchmarks run single-threaded on an Intel Core~i7-10700K CPU ($16$~MiB L3, Windows~11) using release-mode RustPower~\cite{rustpower}, SuiteSparse KLU~\cite{davis2010klu}, and Python~3.13.9 across six MATPOWER and PEGASE test systems~\cite{zimmerman2011matpower,fliscounakis2013} and five RVC test scenarios configured on IEEE~118. Table~\ref{tab:solvers} summarizes the four compared solvers (with pandapower~\cite{thurner2018pandapower} included as a Python parity reference rather than a compiled-speed peer); convergence tolerances are matched at $\tau = 10^{-8}$~p.u.\ for local $Q_{\lim}$ and $\tau = 10^{-10}$~p.u.\ for RVC (pandapower outer tolerance $10^{-6}$~p.u.).

Because outer-loop solvers do not evaluate $\Phi_g$ internally, every converged voltage $\Vbus$ (with net reactive output $Q_g(\Vbus) = \operatorname{Im}(V_g\operatorname{conj}(\Ybus\Vbus)_g) - Q_{\mathrm{fixed},g}$) is audited post-solve via the unscaled ($\gamma_g=1$) KKT residual
\begin{equation}
\begin{aligned}
    \epsilon_{\mathrm{KKT},g}(\Vbus) := \bigl|\,\midop\bigl(&Q_g(\Vbus) - Q_{\max,g},\;\; |V_g| - V_{\mathrm{set},g},\\[-2pt]
    &Q_g(\Vbus) - Q_{\min,g}\bigr)\,\bigr| \le \tau_{\mathrm{KKT}},
\end{aligned}
    \label{eq:kkt_audit}
\end{equation}
which vanishes if and only if~\eqref{eq:local_kkt} holds (simultaneously verifying $Q_{\min,g}\le Q_g\le Q_{\max,g}$, $|V_g|=V_{\mathrm{set},g}$ when unsaturated, and no wrong-sign clamping at $Q_{\max,g}$ or $Q_{\min,g}$). Any generator $g \in \mathcal{G}_{\mathrm{PV}}$ exceeding $\tau_{\mathrm{KKT}} = 10^{-5}$~p.u.\ is counted as a KKT violation in Table~\ref{tab:suite1_qlim}.

\begin{table*}[!t]
\centering
\caption{Suite 2: Coordinated Multi-Generator RVC + Simultaneous Local $Q_{\lim}$ on IEEE~118 ($\tau = 10^{-10}$~p.u.; pp Outer $\tau = 10^{-6}$)}
\label{tab:suite2_rvc}
\setlength{\tabcolsep}{4.0pt}
\renewcommand{\arraystretch}{1.08}
\begin{tabular}{@{}lrrrrrrrr@{}}
\toprule
\textbf{Scenario} & \textbf{RustPower} & \textbf{NR} & \textbf{ls2g 1.3.1 C++} & \textbf{Sym.\ Speedup} & \textbf{ls2g} & \textbf{pandapower} & \textbf{$\max|\Delta V_{\mathrm{tgt}}|$} & \textbf{$\max|\Delta Q|$} \\
 & \textbf{W / C (ms)} & \textbf{It.} & \textbf{W / C (ms)} & \textbf{W/W \;|\; C/C} & \textbf{It.} & \textbf{BinSearch (ms)} & \textbf{vs.\ ls2g (p.u.)} & \textbf{vs.\ ls2g (Mvar)} \\
\midrule
RVC1 (Single unsat.)       & $\mathbf{0.161} / \mathbf{0.295}$ & 6 & $0.427 / 0.712$ & $\mathbf{2.7\times} / \mathbf{2.4\times}$ & 10 & 353.7 & $1.1\times 10^{-16}$ & $1.1\times 10^{-12}$ \\
RVC2 (Dual 70/30 unsat.)   & $\mathbf{0.158} / \mathbf{0.297}$ & 6 & $0.424 / 0.717$ & $\mathbf{2.7\times} / \mathbf{2.4\times}$ & 10 & 531.9 & $0.0$                & $9.0\times 10^{-13}$ \\
RVC3 (Partial sat.\ 8/26)  & $\mathbf{0.135} / \mathbf{0.264}$ & 5 & $0.327 / 0.683$ & $\mathbf{2.4\times} / \mathbf{2.6\times}$ &  7 & 289.8 & $0.0$                & $1.9\times 10^{-11}$ \\
RVC4 (Full sat.\ release)  & $\mathbf{0.127} / \mathbf{0.264}$ & 5 & $0.358 / 0.650$ & $\mathbf{2.8\times} / \mathbf{2.5\times}$ &  8 & 241.6 & $2.2\times 10^{-16}$ & $0.0$                \\
RVC5 (2 coupled groups)    & $\mathbf{0.178} / \mathbf{0.316}$ & 6 & $0.511 / 0.811$ & $\mathbf{2.9\times} / \mathbf{2.6\times}$ & 12 & 912.0 & $1.1\times 10^{-16}$ & $8.9\times 10^{-13}$ \\
\bottomrule
\end{tabular}
\end{table*}

\begin{figure*}[!t]
    \centering
    \includegraphics[width=0.88\textwidth]{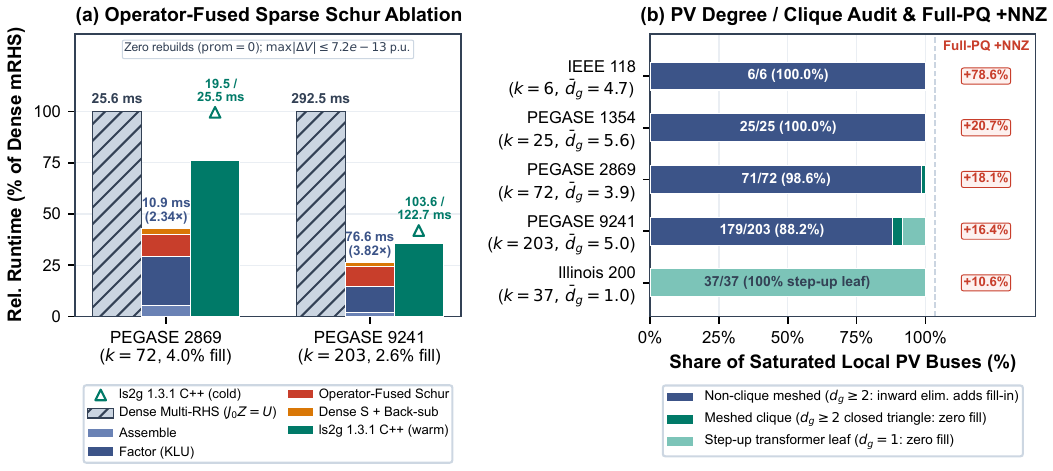}
    \caption{Operator-fused sparse Schur ablation and PV topology audit. (a)~Operator-fused sparse Schur vs.\ cache-blocked dense multi-RHS Schur and C++ lightsim2grid 1.3.1 on PEGASE~2869 ($2.34\times$) and PEGASE~9241 ($3.82\times$, parity $\le 7.2\times 10^{-13}$~p.u.). (b)~Nodal degree ($\bar{d}_g$), Clique-Closure breakdown, and static Full-PQ CSC nonzero expansion ($\mathrm{nnz}(J_{\mathrm{FullPQ}})/\mathrm{nnz}(J_0)-1$) across all five grids.}
    \label{fig:cache_bloat}
\end{figure*}

\begin{figure*}[!t]
    \centering
    \includegraphics[width=0.88\textwidth]{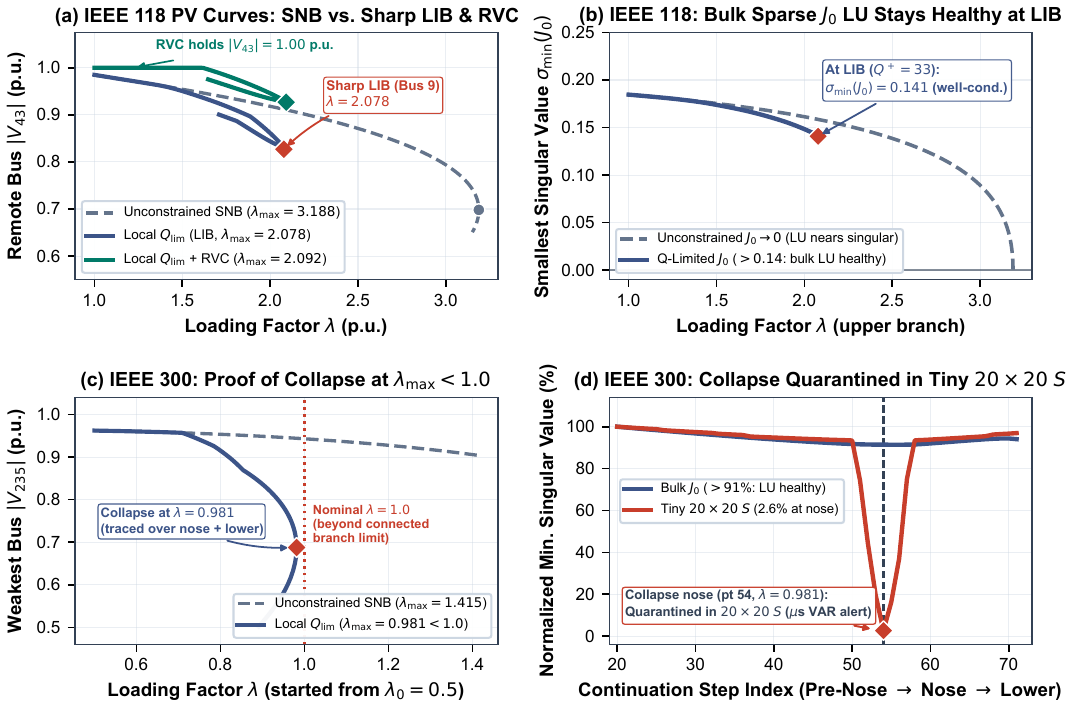}
    \caption{Unified $(k+1)\times(k+1)$ CPF validating Proposition~\ref{prop:spectral_cpf}. (a)--(b)~IEEE~118 unconstrained SNB ($\lambda_{\max}=3.188$), sharp Bus~9 LIB ($\lambda_{\mathrm{LIB}}=2.078$), and RVC plateau raising $\lambda_{\mathrm{LIB}}$ to $2.092$, with bulk $\sigma_{\min}(J_0)\ge 0.141$. (c)--(d)~IEEE~300 traced from $\lambda_0=0.5$: the connected Q-limited branch folds at $\lambda_{\max}=0.981 < 1.0$, keeping $530\times 530$ $J_0$ above $91\%$ conditioning while concentrating fold singularity inside $20\times 20$ $S$ ($\sigma_{\min}(S)\to 2.6\%$).}
    \label{fig:cpf_snb_lib}
\end{figure*}

\subsection{Suite 1: Local Bidirectional $Q_{\lim}$ Accuracy and Speed}
\label{sec:exp_qlim}
Table~\ref{tab:suite1_qlim} summarizes Suite~1. Five grids converge from flat start, whereas on IEEE~300 the Q-limited operating branch folds before nominal load ($\lambda_{\max}=0.981<1.0$, Section~\ref{sec:exp_cpf}). Three results stand out:
\begin{itemize}
    \item \textbf{Bidirectional KKT fidelity:} On PEGASE~9241, one-way pandapower and Legacy QLim over-clamp Bus~6752 at $Q_{\min}=-14.02$~Mvar with $|V|=1.045745 < V_{\mathrm{set}}$ ($\epsilon_{\mathrm{KKT},6752} = 7.86\times 10^{-5} > \tau_{\mathrm{KKT}}$ in~\eqref{eq:kkt_audit}). Both RustPower and lightsim2grid 1.3.1 incur zero KKT violations across all five grids ($\max_g \epsilon_{\mathrm{KKT},g} \le 1.2\times 10^{-10}$~p.u., matching $>12$ digits on Bus~6752).
    \item \textbf{Single-loop SSN vs.\ outer-loop restarts and \texttt{klu\_rgrowth}:} lightsim2grid 1.3.1 restarts NR across 1--3 outer rounds (7--15 iterations) and spends $\approx 30$~ms on PEGASE~9241 running \texttt{klu\_rgrowth} after each \texttt{refactor} to guard fixed pivots under PV$\leftrightarrow$PQ row overwrites. Keeping smooth equations in $J_0$ and partial pivoting in $S$, RustPower converges in 5--9 iterations with zero rebuilds, winning all five grids in cold timing and four of five in warm timing (up to $2.00\times$ on PEGASE~1354, $1.78\times$ on PEGASE~2869, and $76.58 / 102.65$~ms vs.\ $103.63 / 122.67$~ms on PEGASE~9241), and beating Legacy rebuild by $3.3\times$--$20.9\times$.
    \item \textbf{Pass-through at $k = 0$, step-up leaves, and high-$k$ crossover:} When no limits bind ($k = 0$, typical of normal operation and most contingencies),~\eqref{eq:bordered_compact} executes pure $J_0$ solves with zero extra linear-algebra overhead beyond the scalar $Q_g$ bound check, whereas static Full-PQ permanently incurs $+10.6\%$--$+78.6\%$ CSC NNZ expansion (Fig.~\ref{fig:cache_bloat}(b)) and \texttt{klu\_rgrowth} checks. Static Full-PQ is most effective under whole-grid leaf saturation such as Illinois~200, where explicit step-up transformers make all $37/37$ PV buses degree-1 leaves ($\bar{d}_g=1.0$, $+10.6\%$ NNZ) and all 37 saturate simultaneously ($0.92$ vs.\ $1.03$~ms warm; $1.27$ vs.\ $1.15$~ms cold); if desired, a solver can switch to a pre-allocated fallback matrix whenever $k$ exceeds a threshold $k_{\mathrm{th}}$. On meshed grids ($\bar{d}_g=3.9$--$5.6$) and runtime RVC/CPF extensions, active-port Schur wins in both speed and flexibility.
\end{itemize}

\subsection{Suite 2: Coordinated Multi-Generator RVC + Local $Q_{\lim}$}
\label{sec:exp_rvc}
Because standard MATPOWER cases omit remote-bus fields, Suite~2 configures five RVC test scenarios on IEEE~118 alongside local $Q_{\lim}$ (visualized in Fig.~\ref{fig:benchmarks}) to exercise every operating regime: single-unit regulation (RVC1: $42\to 43$), 70/30 sharing (RVC2: $42,44\to 43$), partial saturation with redistribution (RVC3: Bus~42 clamped at $8$~Mvar, Bus~44 absorbing $26.17$~Mvar to hold $|V_{43}|=1.005$~p.u.), full saturation release (RVC4: both clamped at $5$~Mvar, releasing $|V_{43}|=0.9908$~p.u.\ below $V_{\mathrm{set}}=1.030$~p.u.), and two coupled groups (RVC5: $42,44\to 43$ at 70/30 and $19,21\to 20$ at 60/40). Table~\ref{tab:suite2_rvc} and Fig.~\ref{fig:benchmarks}(b) show convergence in 5--6 SSN steps ($<10^{-13}$~p.u.), matching lightsim2grid 1.3.1 to $\le 2.2\times 10^{-16}$~p.u.\ on $|V_{\mathrm{tgt}}|$ while achieving $2.4\times$--$2.9\times$ warm ($2.4\times$--$2.6\times$ cold) speedups (whereas Python pandapower requires $242$--$912$~ms, outside a comparable runtime regime).

\subsection{Suite 3: Sparse Schur Construction Ablation and Clique Audit}
\label{sec:exp_cache}
Across PEGASE~1354, 2869, and 9241, $(X, Y)$ in~\eqref{eq:dual_sparse_xy} contain only $4.9\%$, $4.0\%$, and $2.6\%$ nonzeros ($95.1\%$--$97.4\%$ entrywise zeros), cutting the 9{,}241-bus working set from $27.6$~MB (dense $Z$) to $<300$~KB wavefront slots plus $\approx 1.1$~MB of $Y$ triplets. In Fig.~\ref{fig:cache_bloat}(a), wavefront fusion accelerates blocked dense multi-RHS by $1.27\times$ on Illinois~200 ($1.31\to 1.03$~ms), $1.46\times$ on PEGASE~1354 ($4.09\to 2.81$~ms), $2.34\times$ on PEGASE~2869 ($25.57\to 10.93$~ms), and $3.82\times$ on PEGASE~9241 ($292.52\to 76.58$~ms; $4.73\times$ vs.\ unblocked $362.3$~ms, resolve cut from $290.1$ to $28.1$~ms, $\max|\Delta V|=7.15\times 10^{-13}$~p.u.). Fig.~\ref{fig:cache_bloat}(b) confirms Remark~\ref{rem:clique_closure}: unlike Illinois~200 where explicit step-up transformers make all 37 saturated PV buses degree-1 leaves ($\bar{d}_g=1.0$), saturated PV buses in high-voltage grids sit on meshed substations ($\bar{d}_g=3.9$--$5.6$); aside from occasional $d_g\ge 2$ triangle cliques ($1$ on PEGASE~2869; $7$ triangles + $17$ leaves on PEGASE~9241), $88.2\%$--$100\%$ are non-clique nodes where inward elimination breaks CSC invariance. Although the Schur formulation can incur higher overhead than static pre-allocation on a CPU under extreme leaf saturation, it preserves a uniform base sparsity pattern across batched solves. Because batched GPU $N{-}1$ screening is bottlenecked by divergent sparsity patterns and sequential LU dependencies, sharing a fixed $J_0$ structure while handling active controls in small dense Schur blocks may offer an even greater advantage on GPU architectures.

\subsection{Suite 4: Unified $(k+1)\times(k+1)$ CPF and Bottleneck Diagnosis}
\label{sec:exp_cpf}
Fig.~\ref{fig:cpf_snb_lib} illustrates Proposition~\ref{prop:spectral_cpf} and $J_0$ reuse on IEEE~118 and IEEE~300:
\begin{enumerate}
    \item \textbf{IEEE~118 SNB, sharp LIB, and RVC (Fig.~\ref{fig:cpf_snb_lib}(a)--(b)):} At $k=0$, continuation matches standalone CPF ($\lambda_{\max}=3.18756$). Local $Q_{\lim}$ triggers a sharp Bus~9 LIB at $\lambda_{\mathrm{LIB}}=2.07815$, turned smoothly onto the lower branch without rollbacks; adding RVC ($42,44\to 43$) holds $|V_{43}|=1.000$~p.u.\ over $\lambda\in[1.0,1.62]$ and extends $\lambda_{\mathrm{LIB}}$ to $2.09155$, while $\sigma_{\min}(J_0)\ge 0.141$ stays well-conditioned.
    \item \textbf{IEEE~300 bottleneck diagnosis and feasible boundary (Fig.~\ref{fig:cpf_snb_lib}(c)--(d)):} At $\lambda = 1.0$, unconstrained flow converges whereas all four Q-limited solvers fail. Along the nominal SSN path, bulk $J_0$ stays healthy while the $22\times 22$ Schur matrix $S$ develops a negative real eigenvalue at iteration~6, whose normalized right-eigenvector weight $|\phi_j|^2/\|\phi\|_2^2$ (evaluated in p.u.\ voltage scaling) concentrates $93\%$ on bus~54 (IEEE~300 Bus~63); relaxing that single generator's $Q_{\max}$ restores convergence at $\lambda = 1.0$. Alternatively, early-stopping and switching on the same cached $J_0$ to $(k+1)\times(k+1)$ CPF from $\lambda_0=0.5$ confirms that the connected Q-limited branch folds at $\lambda_{\max}^{\mathrm{qlim}}=0.98139 < 1.0$ (vs.\ $\lambda_{\max}^{\mathrm{un}}=1.41522$, with PV release $Q^+: 13\to 10$ past step~54), keeping $530\times 530$ $J_0$ above $91\%$ conditioning while $\sigma_{\min}(S)$ drops to $2.6\%$ at the nose.
\end{enumerate}

\section{Conclusion}
\label{sec:conclusion}

This paper presented a unified formulation of voltage control constraints for conventional and continuation power flow by combining piecewise-smooth (\texttt{mid}/\texttt{clip}) complementarity equations, \emph{bordered SSN}, and \emph{on-demand Schur reduction} around a fixed, shared base Jacobian $J_0$. Unsaturated PV buses occupy zero border slots, preserving the original NR matrix size at $k = 0$ while enforcing bidirectional local $Q_{\lim}$ and coordinated RVC in a single semismooth Newton loop when $k > 0$. Fusing sparse factor projections directly into the port Schur matrix eliminates intermediate multi-RHS matrices and base Jacobian rebuilds, while appending one continuation port extends the same formulation to CPF with return-path constraint release.

Across SuiteSparse KLU benchmarks up to 9{,}241 buses, the RustPower implementation achieves KKT residuals below $1.2\times 10^{-10}$~p.u., accelerates sparse Schur construction by $3.82\times$ over blocked dense multi-RHS solves, and delivers competitive or faster runtimes than C++ lightsim2grid 1.3.1 (including $2.4\times$--$2.9\times$ speedups on coordinated RVC). On IEEE~300, reusing the same well-conditioned $J_0$ enables both single-snapshot bottleneck diagnosis via $S$ and CPF tracing of the $\lambda_{\max}=0.981 < 1.0$ feasible boundary. Future work will extend this methodology to on-load tap changers (OLTCs), phase-shifting transformers (PSTs), switched shunts, native Rust sparse solvers, and batched GPU contingency screening.

\bibliographystyle{IEEEtran}
\bibliography{bib/qlim_rvc}

\end{document}